\documentclass[10pt,twocolumn,showpacs,preprintnumbers,superscriptaddress,aps,prl]
{revtex4}

\usepackage{graphicx}
\usepackage{amsmath}
\usepackage{color}
\usepackage[usenames,dvipsnames]{xcolor} % just for temp. coloring, can be removed @ end

\begin{document}

\title{Extracting the Mass Dependence and Quantum Numbers of Short-Range Correlated Pairs from $\mathbf{A(e,e^{\prime}p)}$ and $\mathbf{A(e,e^{\prime}pp)}$ Scattering}

\author{C. Colle}
\affiliation{Ghent University, Ghent, Belgium.}
\author{O. Hen}
\affiliation{School of Physics and Astronomy, Tel Aviv University, Tel Aviv 69978, Israel.}
\author{W. Cosyn}
\affiliation{Ghent University, Ghent, Belgium.}
\author{I. Korover}
\affiliation{School of Physics and Astronomy, Tel Aviv University, Tel Aviv 69978, Israel.}
\author{E. Piasetzky}
\affiliation{School of Physics and Astronomy, Tel Aviv University, Tel Aviv 69978, Israel.}
\author{J. Ryckebusch}
\affiliation{Ghent University, Ghent, Belgium.}
\author{L.B. Weinstein}
\affiliation{Old Dominion University, Norfolk VA, USA.}

\begin{abstract}
The nuclear mass dependence of the number of short-range correlated
(SRC) proton-proton (pp) and proton-neutron (pn) pairs in nuclei is a
sensitive probe of the dynamics of short-range pairs in the ground
state of atomic nuclei. This work presents an analysis of
electroinduced single-proton and two-proton knockout measurements off
$^{12}$C, $^{27}$Al, $^{56}$Fe, and $^{208}$Pb in kinematics dominated
by scattering off SRC pairs.  The nuclear mass dependence of the
observed $A(e,e^{\prime}pp)/^{12}\text{C}(e,e^{\prime}pp)$
cross-section ratios and the extracted number of pp- and pn-SRC pairs
are much softer than the mass dependence of the total number of
possible pairs. This is in agreement with a physical picture
of SRC affecting predominantly nucleon-nucleon pairs in a
nodeless relative-$S$ state of the mean-field basis.
\end{abstract}
%25.30.Rw Electroproduction reactions
%25.30.Fj Inelastic electron scattering to continuum
%24.10.-iNuclear reaction models and methods

\maketitle
%
% intro
%
\textit{Introduction:} The nuclear momentum distribution (NMD) is
often quoted as being composed of two separate parts
\cite{Benhar:1994hw,Alvioli:2013ff,Wiringa:2013ala}. Below the Fermi
momentum ($k_F \approx$~250~MeV/c) single nucleons move as independent
particles in a mean field created by their mutual interactions. Above
the Fermi momentum ($k>k_F$) nucleons predominantly belong to
short-range correlated (SRC) pairs with high relative and low
center-of-mass (c.m.) momenta, where high and low are relative to the
Fermi momentum~\cite{Tang:2002ww,Piasetzky:2006ai, Subedi:2008zz,
  Korover:2014dma, Hen:2014}.  In addition to its intrinsic  interest,
the NMD and its division into mean-field and
correlated parts is relevant to two-component Fermi  
systems \cite{Hen:2014lia}, neutrino physics
\cite{PhysRevLett.111.022501,PhysRevLett.111.022502}, and the symmetry energy of nuclear matter \cite{HenSymmetry}.

The mean-field and long-range aspects of nuclear dynamics have
been studied extensively since the dawn of nuclear physics. The effect
of long-range correlations 
%(due primarily to nuclear size effects) 
on the NMDs is limited to  momenta 
which do not extend far beyond $k_F$ \cite{dickhoff2004self}. 
Study of the short-range aspects of nuclear dynamics has blossomed
with the growing availability of high-energy
high-intensity electron and proton accelerators.  Recent
experiments confirm the predictions that SRC pairs dominate the
high-momentum tails ($k>k_F$) of the
NMDs~\cite{Tang:2002ww,Piasetzky:2006ai,Subedi:2008zz,Korover:2014dma},
accounting for 20-25 \% of the NMD probability density 
\cite{Frankfurt:1993sp,Egiyan:2003vg,Egiyan:2005hs,Fomin:2011ng}. These
high-momentum tails have approximately the same shape for all nuclei \cite{Frankfurt:1993sp,
  Egiyan:2003vg, Egiyan:2005hs, Fomin:2011ng, Alvioli:2013ff,
  Wiringa:2013ala, Hen:2014lia, Vanhalst:styfe}, differing only by scale
factors which can be interpreted as a measure of the relative number
of SRC pairs in the different nuclei. In this work, we aim at
understanding the underlying dynamics which give rise to this
universal behavior of the high-momentum tail.

An intuitive picture describing the dynamics of nuclei including SRCs
is that of independent bound nucleons moving in the nucleus,
occasionally getting sufficiently close to each other to temporarily
fluctuate into SRC-induced nucleon-nucleon pairs. This picture can be
formally implemented in a framework in which one shifts the complexity
of the nuclear SRC from the wave functions to the operators by
calculating independent-particle model (IPM) Slater determinant wave
functions and acting on them with correlation operators to include the
effect of SRCs \cite{Roth:2010bm,Vanhalst:styfe,Bogner:2012zm}.  The
observed number of proton-proton (pp) and proton-neutron (pn) SRC
pairs in various nuclei can then be used to constrain the amount and
the quantum numbers of the initial-state IPM nucleon-nucleon
(SRC-prone) pairs that can fluctuate dynamically into SRC pairs
through the action of correlation operators. 

In this paper, we will extract the relative number of pp-SRC and
pn-SRC pairs in different nuclei from measurements of electroinduced
two-proton and one-proton knockout cross-section ratios for medium and
heavy nuclei ($^{27}$Al, $^{56}$Fe, and $^{208}$Pb) relative to
$^{12}$C in kinematics dominated by scattering off SRC
pairs~\cite{Hen:2012jn, Hen:2014}.  In these kinematics in the
plane-wave approximation $A(e,e^{\prime}pp)$ cross sections are
proportional to the number of pp-pairs in the nucleus and
$A(e,e^{\prime}p)$ cross sections are proportional to twice the number
of pp pairs plus the number of pn pairs (2pp+pn).  Therefore, after
correcting the measured cross sections for rescattering of the
outgoing nucleons from the residual nucleus (final state interactions
or FSI), the relative number of pp and pn pairs will be extracted from
measurements of $A(e,e^{\prime}pp)$/$^{12}\text{C}(e,e^{\prime}pp)$
and $A(e,e^{\prime}p)$/$^{12}\text{C}(e,e^{\prime}p)$ cross-section
ratios ~\cite{Hen:2014}.

We will then compare the
$A(e,e^{\prime}pp)$/$^{12}\text{C}(e,e^{\prime}pp)$ cross-section
ratios and the extracted number of pp and pn pairs to factorized
calculations using different models of nucleon pairs in order to
deduce the quantum numbers of the IPM SRC-prone pairs.  We will
provide strong evidence that the relative quantum numbers of the
majority of the SRC-susceptible pairs are $^1 \text{S}_0 (1)$ for pp
and $^3 \text{S}_1(0$) for pn. Hereby, we used the notation
$^{2J+1}\text{L}_S(T)$ to identify the pair's quantum state ($T$ is
the total isospin).

%
% description of the experiment
%
\textit{Experiment:} The one- and two-proton knockout measurements
analyzed in this paper were described in \cite{Hen:2014} and its
supplemental information.  They were carried out using the CEBAF Large
Acceptance Spectrometer (CLAS)~\cite{Mecking:2003zu}, located in
Hall-B of the Thomas Jefferson National Accelerator Facility
(Jefferson Lab) in Newport News, Virginia. The data were collected in
2004 using a 5.014~GeV electron beam incident on $^{12}$C, $^{27}$Al,
$^{56}$Fe and $^{208}$Pb targets. The scattered electron and knocked
out proton(s) were measured with CLAS.  We selected $A(e,e^{\prime}p)$
events in which the electron interacts with a single fast proton from
a SRC nucleon-nucleon pair in the nucleus by requiring large
four-momentum transfer ($Q^2>1.5$ GeV$^2$), Bjorken scaling parameter
$x_B = \frac {Q^2} {2 m_N \omega} > 1.2$ and missing momentum {$300 <
  \vert \vec{p}_{\text{miss}} \vert < 600$~MeV/c}. The four-momentum
transfer $Q^2=\vec{q} \cdot \vec{q} - \left( \frac {\omega}{c} \right)
^2$ where $\vec{q}$ and $\omega$ are the three-momentum and energy
transferred to the nucleus respectively; $m_N$ is the nucleon mass;
the missing momentum $\vec{p}_{\text{miss}} = \vec{p}_p - \vec{q}$,
and $\vec{p}_p$ is the knockout proton three-momentum. We also
required that the knockout proton was detected within a cone of
25$^{\circ}$ of the momentum transfer $\vec{q}$ and that it carried at
least 60\% of its momentum (i.e.~$\frac{\mid \vec{p}_p \mid}{\mid
  \vec{q} \mid} \ge 0.6$). To suppress contributions from inelastic
excitations of the struck nucleon we limited the reconstructed missing
mass of the two-nucleon system $m_{\text{miss}} < 1.1$~GeV/c$^2$. 

\begin{figure}
	\includegraphics[width=0.5\textwidth]{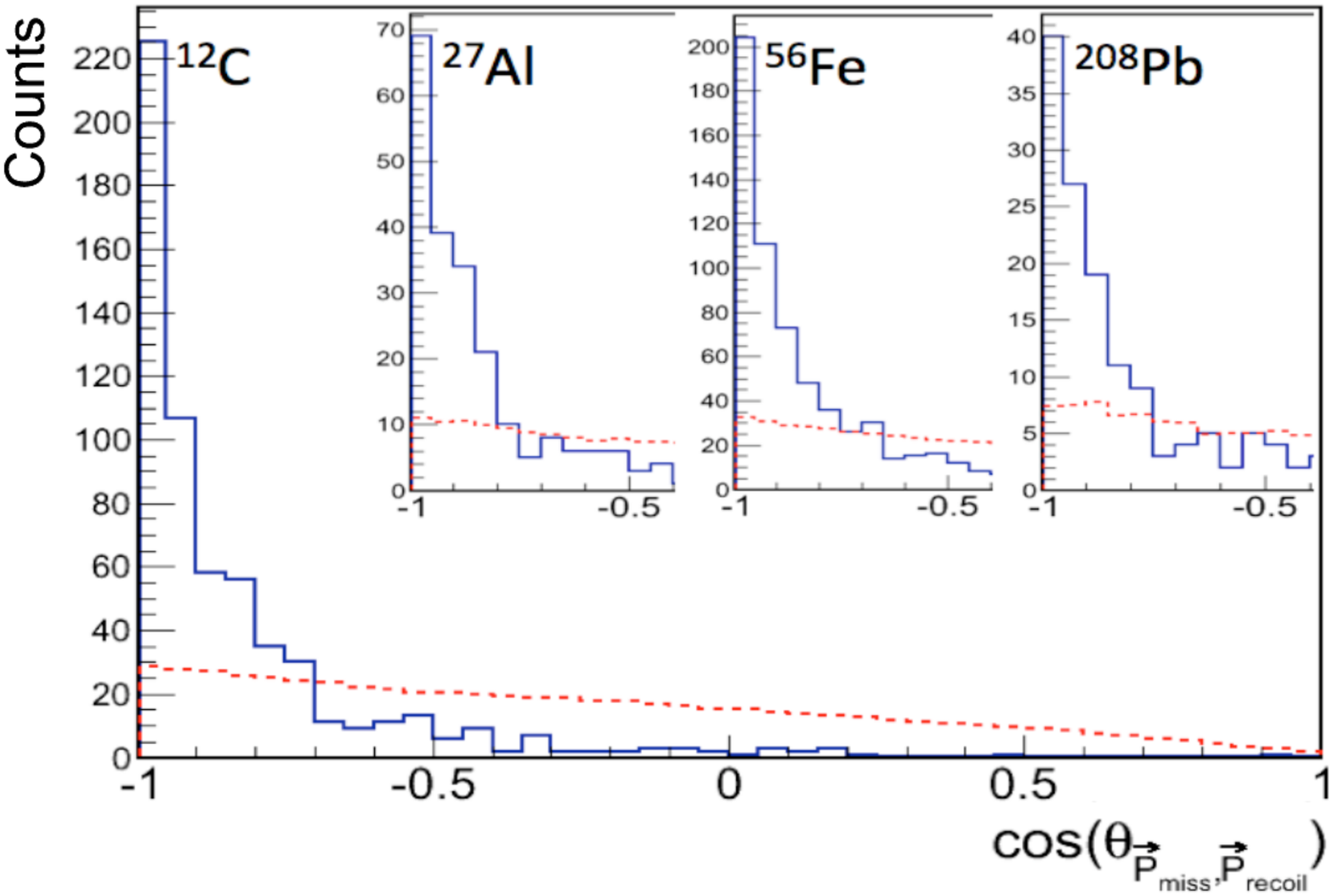}
    \caption{(color online). The distribution of the cosine of the angle between the missing momentum of the leading proton and the recoil proton for $^{12}$C. The insert shows the same distribution for $^{27}$Al, $^{56}$Fe, and $^{208}$Pb. The dashed (red online) line shows the distribution of the random phase-space extracted from mixed-events.} 
    \label{fig:openingAngle}
\end{figure}

The $A(e,e^{\prime}pp)$ event sample contains all $A(e,e^{\prime}p)$
events in which a second, recoil, proton was detected with momentum
greater than 350~MeV/c. Fig.~\ref{fig:openingAngle} shows the
distribution of the cosine of the angle between the initial momentum
of the knockout proton and the recoil proton for these
events~\cite{Hen:2014}.  The recoil proton is emitted almost
diametrically opposite to the missing-momentum direction. The observed
backward-peaked angular distributions are very similar for all nuclei
and are not due to acceptance effects as shown by the angular
distribution of mixed events.
These distributions are a signature of scattering on a nucleon in a SRC pair,
indicating that the two emitted protons were largely back-to-back in
the initial state, having large relative momentum and small
c.m.~momentum~\cite{Subedi:2008zz,Colle:2014}.  Further evidence of
scattering on a SRC nucleon pair is that the recoil proton was emitted
at forward angles (i.e., angles in the range 20-60$^\circ$ with respect
to $\vec q$).

The $A(e,e^{\prime}p)/^{12}\text{C}(e,e^{\prime}p)$ and
$A(e,e^{\prime}pp)/^{12}\text{C}(e,e^{\prime}pp)$ cross-section ratios
are obtained from the ratio of the measured number of events,
normalized by the incident integrated electron flux and the nuclear
density of each target. During the experiment all solid targets were
held in the same location, the detector instantaneous rate was kept
constant, and the kinematics of the measured events from all target
nuclei were almost identical~\cite{Hen:2012jn, Hen:2014}. Therefore
detector acceptance effects cancel almost entirely in the
$A(e,e^{\prime}pp)/\text{C}(e,e^{\prime}pp)$ cross section ratios. Due
to the large acceptance of CLAS, radiative effects affect mainly the
electron kinematics. These corrections were calculated in
Ref.~\cite{Hen:2012jn} for the extraction of the
$A(e,e^{\prime}p)/\text{C}(e,e^{\prime}p)$ cross section ratio. As the
electron kinematics is the same for the $A(e,e^{\prime}p)$ and
$A(e,e^{\prime}pp)$ reactions, the same corrections are used here to
extract the $A(e,e^{\prime}pp)/\text{C}(e,e^{\prime}pp)$ cross-section
ratios. See Ref.~\cite{Hen:2014} for additional details.

%
% FSI RMSGA 
%
\textit{FSI model:} In order to extract the underlying relative number
of pp and pn SRC pairs in nuclei from the measured cross-section
ratios, we must correct the data for FSI effects
~\cite{Hen:2014}. Alternatively, in order to compare the measured
ratios to calculations, we must correct either the data or the
calculation for FSI effects.  The two dominant contributions are: (1)
attenuation of the outgoing nucleon(s) upon traversing the residual
$A-1$ or $A-2$ nucleus, and (2) rescattering of a neutron into a
proton (i.e., single charge-exchange (SCX)). SCX can lead to a pp final state which originates from a pn pair.

The effect of FSIs of the ejected pair with the remaining $A-2$
spectators was computed in a relativistic multiple-scattering Glauber
approximation (RMSGA) \cite{Ryckebusch:2003fc,Cosyn:2013qe}.  The
RMSGA is a multiple-scattering formalism based on the eikonal
approximation with spin-independent NN interactions.  We have included
both the elastic and the SCX rescattering of the outgoing nucleons
with the $A-2$ spectators.  The three parameters entering in the RMSGA
model are taken from $NN$ scattering data and yield an excellent
description of the world's A$(e,e'p)$ transparency data
\cite{Cosyn:2013qe}. The RMSGA caclulations yield attenuation
coefficients that are similar to the power-law results obtained in
analysis of nuclear transparency measurements~\cite{Hen:2012jn}. In
this work no free parameters are tuned to model the FSI effects in the
$A(e,e'p)$ and $A(e,e'pp)$ data under study.

%
% SCX part
%

The SCX probabilities are calculated in a semi-classical
approximation. The probability of charge-exchange re-scattering for a
nucleon with initial IPM quantum numbers $\alpha$ which is brought in
a continuum state at the coordinate $\vec{r}$ is modeled by,
\begin{align}
	 P_{\textrm{CX}}^{\alpha (\beta)}(\vec{r} \; ) = 1 - \exp[ -
           \sigma_{\textrm{CX}}(s) \; \int_{z}^{+\infty} \textrm{d}z'
           \rho^{\alpha \beta}(z')] \, .
     \label{eq:CX_prob_r}
\end{align}
The $z$-axis is chosen along the direction of propagation of the
nucleon with initial quantum numbers $\alpha$. The quantum numbers of
the correlated partner in the SRC pair are denoted with $\beta$. The
$\rho^{\alpha \beta}$ is the density of the residual nucleus available
for SCX reactions. Obviously, for an ejected proton (neutron) only the
neutron (proton) density of the residual nucleus affects SCX
reactions. 
$\sigma_{\textrm{CX}}(s)$ in
Eq.~(\ref{eq:CX_prob_r}), with $s$ the total c.m.~energy squared of
the two nucleons involved in the SCX~\cite{PhysRevC.50.2742}, can be
extracted from elastic proton-neutron scattering
data~\cite{PhysRevC.30.566}.

%
% cross-section model
%
\textit{Cross-section Model:} As outlined in
Refs.~\cite{Ryckebusch:1996wc,Colle:2014}, in the spectator
approximation it is possible to factorize the $A(e,e'pN)$ cross
section in kinematics probing short-range correlated pairs as
\begin{equation}
\frac{\textrm{d} ^{8} \sigma \left[ A (e,e'pN) \right]}{\textrm{d}^{2} \Omega_{e^{\prime}}
\textrm{d}^{3}\vec{P}_{12} \textrm{d}^{3}\vec{k}_{12}} = K_{epN} \sigma _{epN} (\vec{k}_{12}) F^{pN(D)}_{\text{A}}(\vec{P}_{12}) \,,
\label{eq:eeNNfactorized}
\end{equation}
where $\Omega_{e^{\prime}}$ is the solid angle of the scattered
electron, and $\vec{k}_{12}$ and $\vec{P}_{12}$ are the relative and
c.m. momenta of the nucleon pair that absorbed the virtual-photon. The
$K_{epN}$ is a kinematic factor and $\sigma _{epN} (\vec{k}_{12})$ is
the cross section for virtual-photon absorption on a correlated pN
pair. The $F^{pN(D)}_{\text{A}}(\vec{P}_{12})$ is the distorted
two-body c.m. momentum distribution of the correlated pN pair. In the
limit of vanishing FSIs, it is the conditional c.m.~momentum
distribution of a pN pair with relative $\text{S}_{n=0}$ quantum
numbers. Distortions of $F^{pN(D)}_{\text{A}}(\vec{P}_{12})$ due to
FSI are calculated in the RMSGA.  The factorized
cross-section expression of Eq.~(\ref{eq:eeNNfactorized}) hinges on
the validity of the zero-range approximation (ZRA), which amounts to
putting the relative pair coordinate $\vec{r}_{12}$ to zero. The ZRA works as a
projection operator for selecting the very short-range components of
the IPM relative pair wave functions.

The probability for charge-exchange reactions in pN knockout is
calculated on an event per event basis, using the SRC pair probability
density $F^{pN(D)}_{\text{A}}(\vec{R}_{12})$ in the ZRA corrected for
FSI. With the aid of the factorized cross-section expression of
Eq.~(\ref{eq:eeNNfactorized}), the phase-space integrated
$A(e,e^{\prime}pN)$ to $^{12}\text{C}(e,e^{\prime}pN)$
cross-section ratios can be approximately expressed as integrals over
distorted c.m.~momentum distributions,
%
% cross section ratios -> integrated c.m. distr. ratios
%
\begin{multline}
%\frac{\sigma [A(e,e'pN)]}{\sigma[^{12}\text{C}(e,e'pN)]} =
\frac
{\sigma \left[ A(e,e'pN) \right]}
{\sigma \left[ ^{12}\text{C}(e,e'pN) \right]} 
\approx \\
\frac{\int \textrm{d}^{2} \Omega_{e^{\prime}} \textrm{d}^{3} \vec{k}_{12}
K_{{epN}} \sigma_{{epN}}(\vec{k}_{12}) \int \textrm{d}^{3}
\vec{P}_{12} F^{pN(D)}_{\textrm{A}}(\vec{P}_{12})}{\int \textrm{d}^{2}
\Omega_{e^{\prime}} \textrm{d}^{3}\vec{k}_{12} K_{{epN}}
\sigma_{{epN}}(\vec{k}_{12})\int \textrm{d}^{3} \vec{P}_{12}
F^{pN(D)}_{\textrm{C}}(\vec{P}_{12})} \\
= \frac{\int \textrm{d}^{3} \vec{P}_{12} F^{pN(D)}_{\textrm{A}}(\vec{P}_{12})}{\int
\textrm{d}^{3} \vec{P}_{12} F^{pN(D)}_{\textrm{C}}(\vec{P}_{12})} \, .
\label{eq:pN_ratios}
\end{multline}
In the absence of FSI, the integrated c.m. momentum distributions
$\int \textrm{d}^{3} \vec{P}_{12}
F^{pN(D)}_{\textrm{A}}(\vec{P}_{12})$ equal the total number of
SRC-prone pN pairs in the nucleus $A$. Hence, the cross section
ratios of Eq.~(\ref{eq:pN_ratios}) provide access to the relative
number of SRC pN-pairs up to corrections stemming from FSI.
%
% phase space sampling/extrapolation part
%
We have evaluated the ratios of the distorted c.m. momentum
distributions of Eq.~(\ref{eq:pN_ratios}) over the phase space covered
in the experiment. Given the almost $4\pi$ phase space and the high
computational requirement of multidimensional FSI calculations, we use
an importance-sampling approach. The major effect on the c.m.~momentum
distribution $F^{pN(D)}_{\textrm{A}}(\vec{P}_{12})$ when including
FSIs is an overall attenuation, the shape is almost
unaffected~\cite{Colle:2014}. Motivated by this, we used the
c.m. momentum distributions without FSI as sampling distribution for
the importance sampling in the FSI calculations.  When convergence is
reached, the computed impact of FSI is extrapolated to the whole phase
space.

% figure of pp cross section ratios
% SCX INCLUDED
\begin{figure}
	\includegraphics[width=0.5\textwidth]{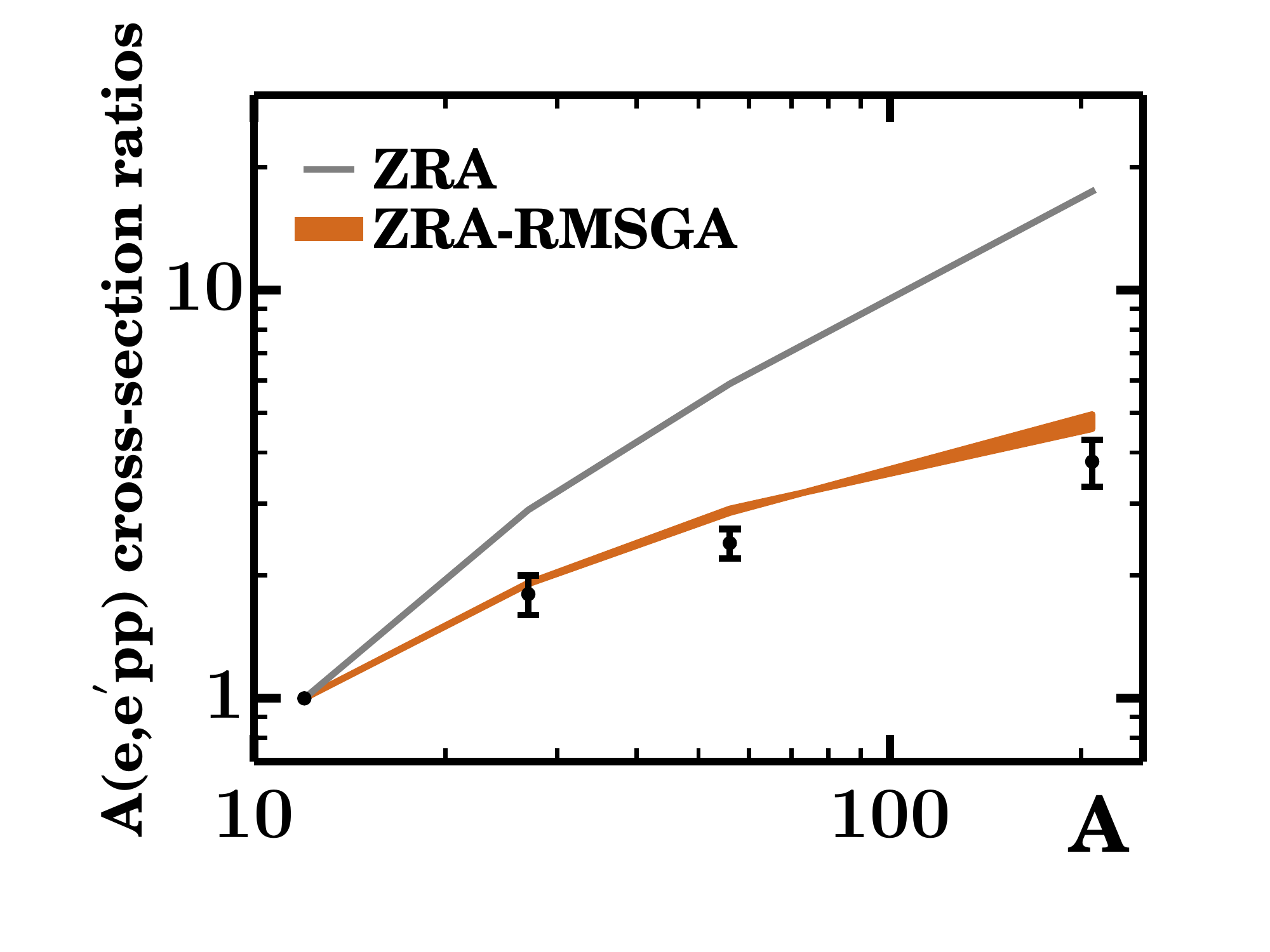}
    \caption{(color online). The mass dependence of the
      $A(e,e^{\prime}pp)/^{12}\text{C}(e,e^{\prime}pp)$
      cross-section ratios. The points show the measured, uncorrected,
      cross section ratios.  The lower orange band  and upper grey
      line denote ZRA reaction-model calculations for $^{12}$C,
      $^{27}$Al, $^{56}$Fe, and $^{208}$Pb based on
      Eq.~(\ref{eq:pN_ratios}) with and without FSI corrections
      respectively. The width of the ZRA-RMSGA band reflects the maximum
      possible effect of SCX.}
    \label{fig:pp_ratios}
\end{figure}

%
% show results
%
\textit{ Results:} Figure~\ref{fig:pp_ratios} shows the measured uncorrected
$\frac{\sigma \left[ A(e,e'pp) \right]} {\sigma \left[
    ^{12}\text{C}(e,e'pp) \right]}$ cross-section ratios compared with
the ZRA reaction-model calculation with and without
RMSGA FSI
corrections. The first striking observation is that the measured 
 cross-section ratios increase very slowly with $A$ (e.g., the Pb/C ratio
is only $3.8\pm 0.5$). For contrast, combinatorial scaling based
on the number of pp pairs leads to a ratio of over 200. The ZRA-RMSGA
calculations agree well with the measured data, yielding a Pb/C ratio
of $4.96^{+0.11}_{-0.14}$. The ZRA and ZRA-RMSGA calculations assume
that only pairs with a finite probability density at relative
coordinate zero contribute to the cross-section. This is
consistent with assuming that only IPM pairs in a relative S$_{n=0}$
state contribute.

\begin{figure}
	\includegraphics[width=0.45\textwidth]{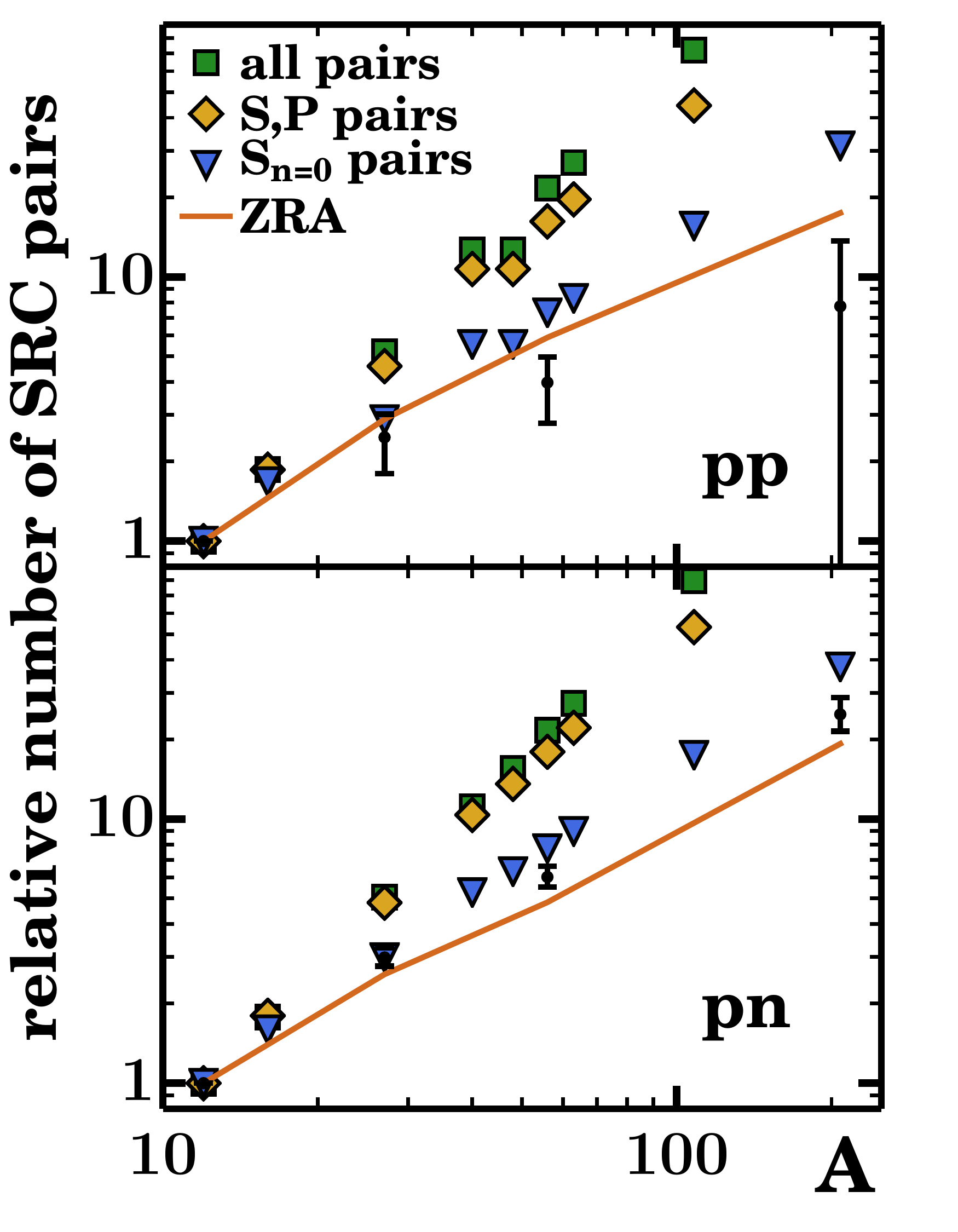}
    \caption{(color online). The mass dependence of the number of pp
      (top panel) and pn (bottom panel) SRC pairs of nucleus A
      relative to $^{12}$C. Data are extracted from the measured CLAS
      $A(e,e'p)$ and $A(e,e'pp)$ cross section
      ratios~\cite{Hen:2012jn, Hen:2014} after correcting for FSI.
      Error bars include the estimated uncertainty on the
      cross-section ratios and the FSI corrections. The green squares
      correspond with unconditional counting of the pp pairs
      i.e.~(Z(Z-1)/30 in the upper panel) and pn pairs (ZN/36 in the
      bottom panel) for the nuclei $^{12}$C,
      $^{16}$O, $^{27}$Al, $^{40}$Ca, $^{48}$Ca, $^{56}$Fe, $^{63}$Cu,
      $^{108}$Ag and $^{208}$Pb. The yellow diamonds are the ratios
      obtained by counting IPM pairs in a relative S and P state. The
      blue triangles count IPM $\text{S}_{n=0}$ pairs. The solid line denotes
      the result of a reaction-model calculation for scattering from
      close-proximity pairs (Eq.~(\ref{eq:pN_ratios})) which takes
      full account of the experimental phase space. This calculation
      does not include FSI corrections as these are applied to the
      data, see text for details.}
    \label{fig:pairs}
\end{figure}

Figure ~\ref{fig:pairs} shows the number of pp- and pn-SRC
pairs in various nuclei relative to Carbon extracted from the measured
$A(e,e'pp)/{\rm C}(e,e'pp)$ and
$A(e,e'p) /{\rm C}(e,e'p)$ cross-section ratios following the method outlined in
Ref. \cite{Hen:2014} with RMSGA corrections for FSI and SCX. The
extracted number of pp pairs are very sensitive to SCX. If the virtual photon is absorbed
on a pn pair and the neutron subsequently undergoes a single charge
exchange reaction with a proton, two protons will be detected in the final
state.  These events must be subtracted in order to extract the number
of pp-SRC pairs. As the contribution from these pn pairs to the pp
final state is comparable to the number of initial pp pairs, this
leads to a large uncertainty in the number of pp pairs, especially for
heavy nuclei.

Figure ~\ref{fig:pairs} also shows the expected number of pp and pn
SRC pairs relative to Carbon for different quantum numbers of the IPM
pairs that can dynamically form SRC pairs through the action of
correlation operators. These include (a) all possible NN pairs
(i.e. $\text{Z(Z-1)}/(6\cdot5)$ and $\text{ZN}/(6\cdot6)$ for pp and
pn pairs respectively), (b) pairs in a nodeless relative S state
(i.e. $\text{S}_{n=0}$), and (c) $\text{L} \le 1$ pairs (i.e.~both S
and P state pairs).   Those "$\text{S}_{n=0}$" pairs are
characterized by the ($n=0,L=0$) quantum numbers for their relative
orbital motion. Of all possible states for the pairs, the
$\text{S}_{n=0}$ pairs have the highest probability for the two
nucleons in the pair to approach each other closely.  Close-proximity
IPM pn pairs in a ${}^3 \text{S}_1(0)$ state are highly susceptible to
the tensor correlation operator that creates SRC pairs in a
spin-triplet state with predominantly deuteron-like quantum numbers
($L=0,2 ; T=0 ; S=1$).

We determine the number of pairs in each case
using an IPM harmonic-oscillator basis and performing a standard
transformation to relative and center-of-mass coordinates as detailed
in Ref.~\cite{Vanhalst:2011es}. The relative number of pairs are
displayed in Fig.~\ref{fig:pairs} and listed in Table~\ref{tab:pairs}.
As can be seen, both (a) the naive combinatorial assumption and (c)
the calculations that include IPM S and P pairs contributions both
drastically overestimate the increase in the number of pairs with $A$.
The ZRA and S$_{n=0}$ pairs counting calculations are in fair
agreement with the extracted number of pp and pn pairs.

As both the ZRA and the $\textrm{S}_{n=0}$ pair counting project IPM
states onto close-range pairs, we expect the two methods to produce a
similar mass dependence of the number of SRC pairs The ZRA predicts a
somewhat softer mass dependence ($\propto \textrm{A}^{1.01 \pm 0.02}$
vs $\textrm{A}^{1.12 \pm 0.02}$). This can be explained by the fact
that the ZRA is a more restrictive projection on close-proximity pairs
than the S$_{n=0}$ counting which accounts also for $\vec{r}_{12}\neq
0$ contributions.

The observed agreement with the experimental data indicates that
correlation operators acting on IPM $\text{S}_{n=0}$ pairs are
responsible for the largest fraction of the high-momentum nucleons in
nuclei. This gives further support to the assumption that the number
of IPM pairs with quantum numbers $\text{S}_{n=0}$ is a good proxy for
the number of correlated pairs in any nucleus $A$
\cite{Vanhalst:styfe,Vanhalst:2011es,Vanhalst:2012ur}.  This is also
consistent with an analysis of the cross section of the ground-state
to ground-state transition in high-resolution
$^{16}\text{O}(e,e^{\prime}pp)^{14}\text{C}$ measurements
\cite{Onderwater:1997zz,Starink2000} which provided evidence for the
$^1 \text{S}_0 (1)$ dominance in SRC-prone pp pairs.

%%%%%
% 2 panel figure of pp and pn pairs
% with full FSI correction on CLAS data
%%%%

%
% Summarizing pair table extraced from CLAS data
%
\begin{table}
\centering
\begin{tabular}{c r r r r r r} \hline \hline
& \multicolumn{3}{c}{pp} & \multicolumn{3}{c}{pn} \\ & S$_{n=0}$ & ZRA
  & \multicolumn{1}{c}{expt.} & S$_{n=0}$ & ZRA &
  \multicolumn{1}{c}{expt.} \\ \hline $^{27}$Al / $^{12}$C & 3.10 &
  2.89 & $2.47^{+0.55}_{-0.67}$ & 2.99 & 2.52 & $
  2.99^{+0.26}_{-0.22}$ \\ $^{56}$Fe / $^{12}$C & 8.60 & 5.89 &
  $3.98^{+0.99}_{-1.19}$ & 7.72 & 4.82 & $ 6.03^{+0.60}_{-0.51}$
  \\ $^{208}$Pb/ $^{12}$C &45.29 & 17.44& $7.73^{+5.92}_{-7.23}$
  &37.62 &18.80 & $24.87^{+3.89}_{-3.42}$ \\ \hline \hline
\end{tabular}
\caption{The relative number of SRC pp and pn pairs calculated using
  S$_{n=0}$ counting and the ZRA reaction model compared to the extracted
  values from the measured $A(e,e'p)$ and $A(e,e'pp)$ ratios after
  correcting for FSI effects. The error includes the uncertainties on
  the cross-section ratios and FSI calculations.}
\label{tab:pairs}
\end{table}

\textit{Conclusions:} We have extracted the relative number of np and
pp SRC correlated pairs in nucleus $A$ relative to Carbon from
previously published measured $A(e,e'pp)/{\rm C}(e,e'pp)$ and
$A(e,e'p)/{\rm C}(e,e'p)$ cross section ratios corrected for final
state interactions.  The relative number of np and pp pairs increases
much more slowly with $A$ than expected from simple combinatorics.

We calculated the cross section in a framework
which shifts the complexity of the nuclear SRC from the wave functions
to the operators by calculating independent-particle model (IPM)
Slater determinant wave functions and acting on them with correlation
operators to include the effect of SRCs
\cite{Roth:2010bm,Vanhalst:styfe,Bogner:2012zm}.  This framework is
also well suited for explaining the recently observed similarities
between the contact term measured in ultra-cold atomic systems and
that extracted for atomic nuclei \cite{Hen:2014lia}.
The uncorrected $A(e,e'pp)/{\rm C}(e,e'pp)$ cross section ratios are
consistent with a zero range approximation (ZRA) calculation including
the effects of FSI.

Due to factorization, the ratio of calculated cross sections is
approximately equal to the ratio of the distorted c.m.~momentum
distributions.  In the absence of FSI, the integrated c.m.~momentum
distribution equals the total number of SRC-prone pairs in that
nucleus.  We compared three choices of SRC-prone pairs to the data:
(a) all pairs, (b) pairs in a nodeless relative S state ($\text{S}_{n=0}$),
and (c) $L\le 1$ pairs (i.e., both $\text{S}$ and $\text{P}$).
 
We found that the soft mass dependence of the measured $A(e,e'pp)$
cross-section ratios agrees with scattering from highly selective
close-proximity pairs (i.e., only IPM relative $\textrm{S}_{n=0}$
pairs). The mass dependence of the extracted ratios of the number of
short-range correlated pp and pn pairs provides additional support for
this conclusion.  All these results consistently hint at a physical
picture whereby the aggregated effect of SRC in the nuclear wave
function is determined to a large extent by mass-independent
correlation operators on $\text{S}_{n=0}$ pairs.

\textit{Acknowledgements:} We acknowledge the efforts of the Jefferson
Lab staff that made this experiment possible and the EG2 group and
CLAS Collaboration.  The Ghent group is supported by the Research Foundation Flanders
(FWO-Flanders) and by the Interuniversity Attraction Poles Programme
P7/12 initiated by the Belgian Science Policy Office. For the
theoretical calculations, the computational resources (Stevin
Supercomputer Infrastructure) and services used in this work were
provided by Ghent University, the Hercules Foundation and the Flemish
Government.  O.~Hen and E.~Piasetzky are supported by the Israeli
Science Foundation.  L.B.~Weinstein is supported by the US Department
of Energy under grant de-SC00006801 and DOE-FG02-96ER40960.

%\bibliography{biblio}

\end{document}